\documentclass[%
 aps,
 prb,%
 amsmath,amssymb,
reprint,
superscriptaddress, 
twocolumn%
]{revtex4-2}

\usepackage{graphicx}
\usepackage{dcolumn}
\usepackage{bm}
\usepackage[version=3]{mhchem} 
\usepackage[T1]{fontenc}       
\usepackage{gensymb}
\usepackage[usenames,dvipsnames]{pstricks}
\usepackage{epsfig}
\usepackage{pst-grad} 
\usepackage{pst-plot}
\usepackage{epstopdf}
\usepackage{soul}
\usepackage{multirow}
\usepackage[breaklinks=true,colorlinks=true,linkcolor=blue,urlcolor=blue,citecolor=blue]{hyperref}
\usepackage{xcolor}

\begin{document}

\preprint{AIP/123-QED}

\title[The submonolayer assembly of \ce{C60F48} surface acceptors on the hydrogen-terminated (100) diamond surface]{The submonolayer assembly of \ce{C60F48} surface acceptors on the hydrogen-terminated (100) diamond surface}

\author{Alex K. Schenk}
\email{A.Schenk@latrobe.edu.au}
\affiliation{Department of Mathematical and Physical Sciences, School of Computing, Engineering and Mathematical Sciences, La Trobe University, Victoria 3086, Australia}

\author{Kevin J. Rietwyk}
\affiliation{School of Science, RMIT University, Melbourne, Vic 3001, Australia}

\author{Mark T. Edmonds}
\affiliation{School of Physics and Astronomy and Monash Centre for Atomically Thin Materials, Clayton Victoria 3800, Australia}

\author{Anton Tadich}
\affiliation{Department of Mathematical and Physical Sciences, School of Computing, Engineering and Mathematical Sciences, La Trobe University, Victoria 3086, Australia}
\affiliation{Australian Synchrotron, 800 Blackburn Road, Clayton, Victoria 3168, Australia}

\author{Chris I. Pakes}
\affiliation{Department of Mathematical and Physical Sciences, School of Computing, Engineering and Mathematical Sciences, La Trobe University, Victoria 3086, Australia}

\begin{abstract}
We have characterised the submonolayer assembly of \ce{C60F48} molecular surface acceptors on the hydrogen terminated (100) diamond surface using scanning tunneling microscopy and X-ray photoelectron spectroscopy, and explored the influence this assembly has on the energy level alignment which governs surface transfer doping. Following deposition, disordered multilayered islands are observed. Annealing to 90\celsius\, causes a transition from dewetting multi-layered islands to a single wetting layer at sufficiently high coverage, accompanied by a change in the energy alignment between the \ce{C60F48} and the diamond surface which consequently decreased the carrier density induced by transfer doping by 20\%. 
\end{abstract}

\maketitle

\section{Introduction}
Over the last two decades, hydrogen-terminated diamond surfaces have generated a great deal of interest due to the two-dimensional hole gas (2DHG) which can be induced by exposure to air \cite{Maier_SC} or the adsorption of appropriate high electron affinity acceptor molecules \cite{strobel2004surface,qi2007surface,edmonds2012surface,verona2016comparative,crawford2018thermally}, in a process referred to as surface transfer doping. The resulting p-type surface conductivity has been exploited for the fabrication of transistors \cite{watkins2023surface}, including nanoscale single-hole transistors \cite{banno2002fabrication}, and it has been shown that the 2DHG possesses a strong and tunable spin-orbit coupling \cite{edmonds2014spin,xing2020strong} making it a potential platform for spintronic devices. The use of high electron affinity molecular acceptors  - typically fluorinated organic molecules \cite{qi2007surface,edmonds2012surface} and transition metal oxides \cite{verona2016comparative,crawford2018thermally} - is of technological interest as it offers a pathway to more direct control over the induced carrier density and greater thermal stability \cite{crawford2018thermally} than the water layer which air-induced transfer doping relies on. 

 \begin{figure}
 	\centering
 	\includegraphics{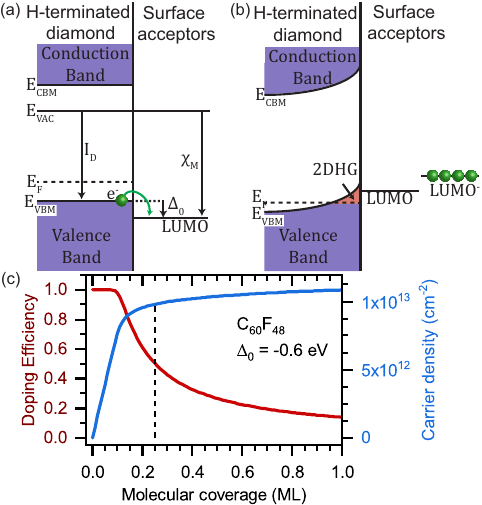}
 	\caption{A summary of the surface-transfer doping process using a high electron affinity molecular dopant, (a) energy level diagram in the limit of vanishing concentration (b) energy level diagram when transfer doping has effectively ceased due to the interface potential, (c) the calculated doping efficiency and carrier density in the 2DHG as a function of \ce{C60F48} coverage, using parameters from Edmonds~\textit{et al.} \cite{edmonds2012surface}. Dashed line indicates the point where the carrier density is 90\% of the value it will reach at 1~ML coverage.}
 	\label{fig:STD_sketch}
 \end{figure}

Surface transfer doping with molecular acceptors can be understood by considering the energy level alignment between the hydrogen-terminated diamond surface and the molecular acceptor, the potential $\Delta\Phi(p)$ generated by charge transfer, and Fermi-Dirac statistics; a detailed mathematical treatment can be found in the work of Edmonds~\textit{et al.} \cite{edmonds2012surface}. In brief, the choice of a molecule with a suitably high electron affinity ($\chi_M\gtrapprox4.5$~eV) combined with the low ionisation potential of hydrogen-terminated diamond ($I_D = 4.2~\rm{eV}$), ensures that when vacuum level alignment occurs the lowest unoccupied molecular orbital (LUMO) of the acceptor is below the diamond valence band maximum (VBM) and Fermi level, as shown in Fig.~\ref{fig:STD_sketch}(a). This alignment is paramaterised in terms of the ``acceptor energy'' of the molecular acceptor, $\Delta_0 = \rm{E_{LUMO} - E_{VBM}}$, in analogy with more conventional semiconductor doping; for an efficient molecular acceptor this quantity is initially negative. This alignment results in spontaneous charge transfer from the surface into the molecular adlayer, and the charge separation produces an interface potential $\Delta\Phi(p)$ dependent on the induced hole sheet density ($p$), typically modelled as a parallel plate capacitor with characteristic capacitance $\rm{C_\square}$ \cite{edmonds2012surface}. As a result, charge transfer results in a downward shift of the Fermi level, and the potential due to charge separation causes the LUMO to shift upward, eventually leading to an energy level alignment in which further charge transfer is no longer possible, as demonstrated in Fig.~\ref{fig:STD_sketch}(b). In practice this means that initially the doping efficiency $\eta$ is unity - one hole is generated in the diamond surface for every molecule on the surface - until a critical coverage is reached, above which there is a probability of molecular acceptors remaining neutral, and the charge transfer process becomes rapidly less efficient. This process is described by the equation:
\[ \eta=\dfrac{{\rm N}^{-}}{{\rm N}_{\rm T}}=\dfrac{1}{\dfrac{1}{g}\exp[(\Delta_{0}+ \dfrac{e\cdot p}{\rm{C_\square}}-w(p))/k_{b}T]+1} \]
where $\rm{N^{-}}$ is the areal density of ionised molecular acceptors, $\rm{N_{T}}$ the total molecular acceptor density, $g$ is the degeneracy of the acceptor LUMO, and $w(p)$ is a universal function for diamond which gives the position of the Fermi level with respect to the diamond valence band maximum (VBM) as a function of hole sheet density $p$ \cite{edmonds2011surface,edmonds2012surface}. As shown in Fig.~\ref{fig:STD_sketch}(c) using \ce{C60F48} as an example, for high electron affinity acceptors $\Delta_0$ is sufficiently large that the critical molecular coverage is submonolayer, and so the doping process is effectively complete at less than one monolayer coverage. 

There is a multitude of published studies characterising molecular acceptors on the hydrogen-terminated diamond surface, typically photoelectron spectroscopy studies of the energy level alignment and doping efficiency of specific acceptors \cite{edmonds2012surface,crawford2016enhanced,russell2013surface}, or electrical transport studies examining the electrical or spin transport properties of the 2DHG which is induced by various acceptors \cite{strobel2004surface,edmonds2014spin,xing2020moo3,xing2020high,xing2020strong}. However, one aspect which has been largely overlooked is the assembly of the molecular acceptor layer and how this influences the transfer doping process. Indeed, to date the only article which considers the assembly of a molecular acceptor on the hydrogen-terminated diamond surface is Nimmrich~\textit{et al.} \cite{nimmrich2012influence}, which used noncontact atomic force microscopy (NC-AFM) to characterise the submonolayer assembly of the relatively weak surface acceptor \ce{C60} \cite{strobel2004surface} on the hydrogen-terminated (100) diamond surface. Nimmrich~\textit{et al.} showed that, contrary to the assumed layer-by-layer (Frank-Van der Merwe) growth  \cite{strobel2004surface}, submonolayer \ce{C60} grows on the surface in single and double layer islands (Volmer-Weber growth), and annealing the film to 505~K lead to the formation of extended double layer islands \cite{nimmrich2012influence}. As single \ce{C60} layer films do not meaningfully participate in transfer doping, while multilayer films do \cite{strobel2004surface}, this transition can be expected to alter characteristics of the 2DHG formed by surface transfer doping with \ce{C60}. As one of the motivations for using molecular acceptors is the thermal stability implicit in the higher sublimation and decomposition temperature of such molecules \cite{crawford2018thermally}, the implicit consequences of the findings of Nimmrich~\textit{et al.} suggest that it may be important to consider whether temperature can influence the assembly and doping of such films, particularly in regards to the behaviour and reproducibility of nanoscale devices.

In this article we explore the submonolayer assembly of \ce{C60F48} on the hydrogen-terminated (100) diamond surface, and the effect annealing to 90\celsius\, has on the assembly and transfer doping, using a combination of Scanning Tunneling Microscopy (STM) and synchrotron-based X-ray Photoelectron Spectroscopy (XPS). In order to examine the impact the neutral molecules have on the assembly we have looked at \ce{C60F48} coverages of 0.05~ML and 0.30~ML; in the former case all molecules are expected to be charged while in the latter case there is a non-negligible density of neutral molecules upon deposition \cite{edmonds2012surface}. Contrary to previous assumptions of layer-by-layer growth, we find that \ce{C60F48} forms disordered, multilayered islands on the surface. Annealing causes no change at 0.05~ML, but at 0.30~ML there is a transition to larger, single layer wetting layers, accompanied by a change in the energy level alignment between the surface and the adlayer and a 20\% reduction in the charge transferred to the adlayer and thus the carrier density of the 2DHG. This is the first investigation of molecular surface transfer doping which combines scanning probe microscopy and synchrotron spectroscopy to examine the interplay between molecular assembly and charge transfer energetics, and demonstrates the importance of considering the influence of molecular assembly when characterising transfer doping and implementing it in real-word devices.

\section{Experimental Methods}
\subsection{Ex situ sample preparation}
Experiments were performed on a synthetic, boron doped (100) single crystal diamond (Element Six) with a boron concentration in the range $5\times10^{18}-5\times10^{19}~\rm{cm}^{-3}$. The sample was cleaned by boiling in an acid mixture (\ce{H2SO4}/\ce{HClO4}/\ce{HNO3}, 1:1:1) to remove metallic contamination and non-diamond carbon phases, followed by hydrogen-termination in a microwave hydrogen plasma at a power of 2~kW for 45 minutes while the sample temperature was maintained at 800\celsius. The sample was exposed to air before being transferred into each ultrahigh vacuum (UHV) system. 

\subsection{STM}
Scanning Tunneling Microscopy experiments were performed in an UHV system made up of two chambers, an analysis chamber equipped with a SPECS Aarhus 150 STM, and a preparation chamber for \textit{in situ} sample preparation. The sample was annealed \textit{in vacuo} to 450\celsius\, for 1 hour to remove atmospheric contamination \cite{schenk2016high}. Measurements were taken prior to adlayer deposition to ensure surface cleanliness. \ce{C60F48} was deposited by sublimation from a quartz crucible, using a quartz crystal monitor (QCM) to determine the deposition rate, guided by rates measured in the XPS experiments with an identical QCM in a similar geometry. All preparation was carried out at a pressure of less than $5\times10^{-8}$~mbar.

All measurements were performed at room temperature, and the base pressure of the analysis chamber was maintained at below $5\times10^{-10}$~mbar during all measurements. The images presented herein have been calibrated using the atomic spacings from STM images measured during this study of a well prepared Ag(111) surface or HOPG surface.
\subsection{XPS}
XPS measurements were conducted at the Soft X-ray Spectroscopy (SXR) beamline at the Australian Synchrotron. The sample was annealed \textit{in vacuo} to 450\celsius\, for 1 hour to remove atmospheric contamination \cite{schenk2016high}. Measurements were taken prior to \ce{C60F48} deposition to ensure the surface was appropriately prepared.  \ce{C60F48} deposition was performed using a Knudsen cell (NTEZ low temperature effusion cell, MBE Komponenten) equipped with a quartz crucible, set to a temperature of 195\celsius. A QCM was used as a guide to the deposition rate, and the coverage was subsequently confirmed using photoelectron attention as in previous work \cite{edmonds2012surface}. The sample was measured at pressures below $5\times10^{-10}$~mbar, and all \textit{in situ} preparation was carried out at pressures less than $5\times10^{-8}$~mbar.

The measured core levels were analysed using a Shirley background subtraction \cite{shirley1972high} and the Voigt lineshape to represent each component, using a fitting model described in detail in the text. A Lorentzian width of 0.15~eV was used to reflect the lifetime of the C1s core hole \cite{Graupner_XPS}. The binding energy scale of all spectra are referenced to a common Fermi level by measuring the $\rm{Au}4f_{7/2}$ peak position of a clean gold foil in electrical contact with the sample, and setting this to a binding energy of 84.00~eV. 

\section{Results and Discussion}
\subsection{Scanning Tunneling Microscopy}

\begin{figure}
	\centering
	\includegraphics{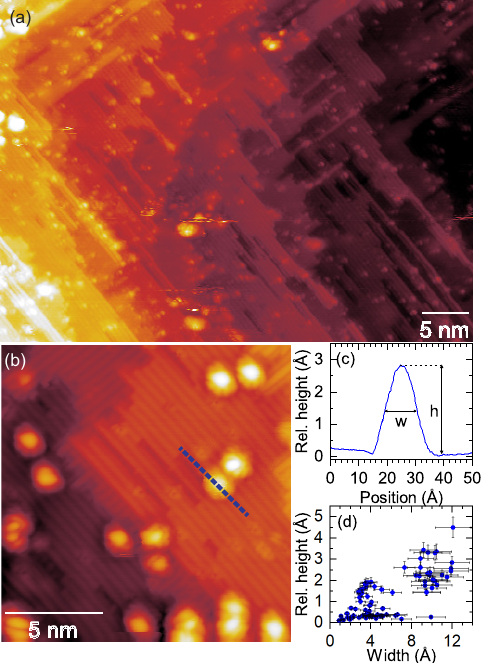}
	\caption{STM images of the as-prepared hydrogen-terminated (100) diamond surface taken with settings (a) scan dimensions $700~\text{\AA}\times500~\text{\AA}$ $V_{t} = -2.360~\rm{V}$,  $I_{t} = -0.640~\rm{nA}$ (b) scan dimensions $150~\text{\AA}\times150~\text{\AA}$ $V_{t} = 3.01~\rm{V}$,  $I_{t} = 1.06~\rm{nA}$. (c) shows a linescan of feature along the line indicated in (b), (d) shows a plot of feature height against feature full width at half maxima for dangling bond and defect features on the diamond surface. These feature dimensions are extracted from multiple images taken on different sample preparations and a range of tip conditions.}
	\label{fig:Hterm_STM}
\end{figure} 

An overview of the as-prepared state of the hydrogen-terminated (100) diamond surface, as imaged by STM, is presented in Figure~\ref{fig:Hterm_STM}. As expected for a high quality surface, large-scale images, such as Fig.~\ref{fig:Hterm_STM}(a), show atomically flat terraces which are typically 5 - 10~nm wide and have irregular edges. These terraces are made up of dimer rows which can be clearly distinguished in higher resolution images, such as Fig.~\ref{fig:Hterm_STM}(b); in adjacent terraces the row direction is rotated by 90\degree. The surface is populated by a low density of small protrusions and atomic-scale irregularities; Figure~\ref{fig:Hterm_STM}(c) shows a linescan for one of the larger protrusions from Fig~\ref{fig:Hterm_STM}(b), and Fig.~\ref{fig:Hterm_STM}(d) shows the dimensions of a representative set of such features, determined from various images of the as-prepared surface acquired throughout this study. Based on previous STM investigations of the diamond surface, we assume the protrusions are dangling bonds \cite{bobrov2003atomic} and the irregularities are surface defects \cite{kuang1995surface,bobrov2003atomic2,nimmrich2010atomic}. These are common features of well-prepared hydrogen-terminated surfaces \cite{ley2011preparation}, and so will be present in all studies of surface transfer doping to date and therefore do not impact any comparison between this and existing work. Importantly, as the majority of these features are less than 3~\AA\, in height, and all are less than the 7 - 10~\AA\, height of a typical \ce{C60}-derived fullerene when measured with STM \cite{rahe2012dewetting,smets2013charge}, we can be confident in differentiating between those features which are an intrinsic part of the diamond surface and individual \ce{C60F48} molecules in later measurements. 

\begin{figure*}
	\centering
	\includegraphics[width=172.5mm]{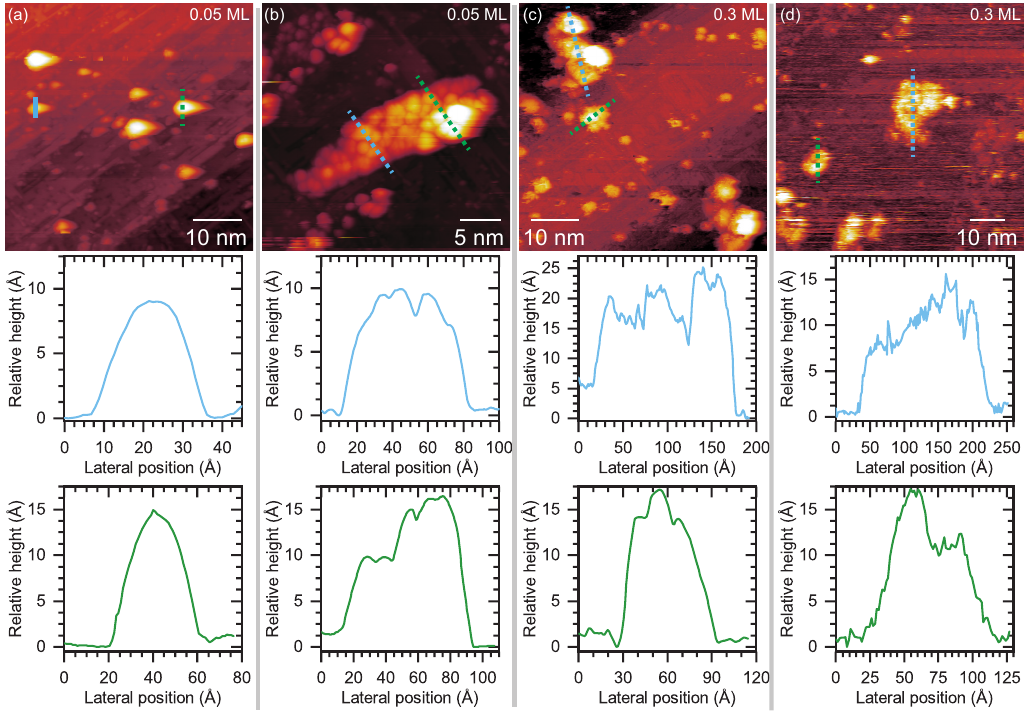}
	\caption{Representative STM images of the \ce{C60F48} layer on the hydrogen-terminated diamond (100) surface following deposition (a) coverage 0.05~ML, scan dimensions $500~\text{\AA}\times500~\text{\AA}$, $V_{t} = -2.75~\rm{V}$,  $I_{t} = -0.18~\rm{nA}$, (b) coverage 0.05~ML, scan dimensions $300~\text{\AA}\times300~\text{\AA}$, $V_{t} = -3.00~\rm{V}$,  $I_{t} = -0.22~\rm{nA}$, (c) coverage 0.30~ML, scan dimensions $500~\text{\AA}\times500~\text{\AA}$, $V_{t} = -0.806~\rm{V}$,  $I_{t} = -0.550~\rm{nA}$, (d) coverage 0.30~ML, scan dimensions $700~\text{\AA}\times700~\text{\AA}$, $V_{t} = -0.806~\rm{V}$,  $I_{t} = -0.660~\rm{nA}$}
	\label{fig:C60F48_deposited_STM}
\end{figure*}
 
A representative set of STM images acquired following the deposition of 0.05~ML and 0.3~ML of \ce{C60F48} is shown in Fig.~\ref{fig:C60F48_deposited_STM}, along with linescans through a selection of features. What we observe following the deposition are irregular features substantially larger - both laterally and vertically - than the features intrinsic to the diamond surface identified above, and which are therefore individual \ce{C60F48} molecules or islands of \ce{C60F48}. We note that in some images, such as Fig~\ref{fig:C60F48_deposited_STM}(a), some of the features elongate in the fast scan direction (left to right) of the raster, which is likely an artefact resulting from the tip strongly interacting with the weakly physisorbed \ce{C60F48} molecules \cite{nawaz1992stm,bohringer1998scanning}. Furthermore in some cases, such as Fig.~\ref{fig:C60F48_deposited_STM}(d), imaging these features is only possible with tunneling conditions which do not allow detailed imaging of the diamond surface details. Regardless, we are able to use this data to discern some key information about the adlayer formation. It has typically been assumed that \ce{C60F48} adlayers grow on the hydrogen-terminated diamond surface in a Frank-van der Merwe (layer-by-layer) mode \cite{strobel2004surface}. If this were the case, then at submonolayer coverage we would expect to see individual isolated molecules or islands of molecules which are a single molecule in height. At 0.05~ML coverage the lateral dimensions of the observed features are typically in the $10 - 70$~\AA\, range, consistent with individual \ce{C60F48} molecules and small islands, although infrequently there are larger islands up to 300~\AA\, in width as shown in Fig~\ref{fig:C60F48_deposited_STM}(b). In the 0.30~ML data we see similar features, although islands up to 160~\AA\, in width are regularly observed. Regardless of the coverage, as the linescans show, the height of these features is inconsistent; the maximum height of each feature ranges from $7.2\pm1.0$~\AA, the expected height of a single \ce{C60F48} molecule \cite{smets2013charge}, up to $26.6\pm2.0$~\AA, in height. As many of these islands are substantially taller than the height of a single molecule, we can infer that many of the islands are made up of multiple molecular layers. As the measured heights are not discrete values, but rather have a broad distribution, and the finer structure of these islands - when observable - is irregular, we believe that these islands are disordered groupings of molecules, similar to the case of \ce{C60F48} on zinc-tetraphenylporphyrin films \cite{smets2013doping}, rather than the ordered islands which \ce{C60} forms on the hydrogen-terminated diamond (100) surface \cite{nimmrich2012influence}.

The observed Volmer-Weber dewetting island growth is unexpected, as the present understanding of transfer doping of hydrogen-terminated diamond with \ce{C60F48} indicates that a large proportion of the molecules will be negatively charged at these submonolayer coverages \cite{edmonds2012surface} - indeed, at 0.05~ML all molecules should be charged - and so should electrostatically repel one another. We infer from the observed growth that the negative charge on \ce{C60F48} molecules participating in transfer doping is screened, such that the electrostatic repulsion is smaller than the attractive intermolecular van der Waals force between \ce{C60F48} molecules, which for fullerenes is typically of the order of 0.2~eV \cite{pan1991heats,gravil1996adsorption} between two molecules. This island formation does, however, provide an explanation for the bulk-like electron affinity calculated for neutral \ce{C60F48} on hydrogen-terminated (100) diamond. Edmonds~\textit{et al.} derived a value of $4.97\pm0.2$~eV from photoelectron spectroscopy measurements \cite{edmonds2012surface}, significantly closer to the 5.0~eV of bulk \ce{C60F48} \cite{mitsumoto1996soft,mitsumoto1998electronic} than to the 4.06~eV of isolated \ce{C60F48} molecules \cite{jin1994attachment}. Lacking evidence of any alternative explanation, the anomalously large electron affinity was attributed to the diamond substrate screening the molecules \cite{edmonds2012surface}; however, it can now be suggested that this is due to the \ce{C60F48} island formation resulting in a bulk-like screening of the neutral molecules. 

One aspect of this assembly which we are unable to assess from the STM images is the distribution of charged and neutral across the surface and within the islands at 0.3~ML, as there is no indication which molecules are charged and which are neutral. Given that the doping efficiency is initially unity and that clearly the charged \ce{C60F48} molecules are capable of nucleating islands, we might expect that when neutral molecules appear they add to these islands, resulting in an inhomogeneous distribution of charged and neutral molecules through the depth of the islands. We will consider this question in section~\ref{sec:XPS}, when characterising the adlayer with photoelectron spectroscopy. 

\begin{figure}
	\centering
	\includegraphics[width = 81mm]{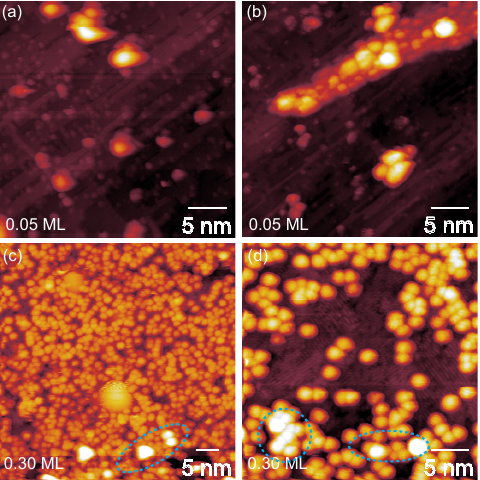}
	\caption{Representative STM images of the \ce{C60F48} layer following annealing to 90\celsius\, (a) coverage 0.05~ML, scan dimensions $300~\text{\AA}\times300~\text{\AA}$,  $V_{t} = -2.02~\rm{V}$,  $I_{t} = -0.21~\rm{nA}$,(b) coverage 0.05~ML, scan dimensions $300~\text{\AA}\times300~\text{\AA}$, $V_{t} = -2.36~\rm{V}$,  $I_{t} = -0.22~\rm{nA}$, (c) coverage 0.30~ML, scan dimensions $500~\text{\AA}\times500~\text{\AA}$, $V_{t} = 4.64~\rm{V}$,  $I_{t} = 1.58~\rm{nA}$, (d) coverage 0.30~ML, scan dimensions $300~\text{\AA}\times300~\text{\AA}$, $V_{t} = 4.70~\rm{V}$,  $I_{t} = 1.55~\rm{nA}$}
	\label{fig:C60F48_annealed_STM}
\end{figure}

Figure~\ref{fig:C60F48_annealed_STM} shows a representative set of STM images acquired after annealing the adlayer at 90\celsius\, for 10 minutes. For the 0.05~ML coverage we still see small islands of molecules over much of the surface (Figure~\ref{fig:C60F48_annealed_STM}(a)), with infrequent observations of larger extended islands (Figure~\ref{fig:C60F48_annealed_STM}(b)); all observed features are less than 25~\AA\, high, relative to the surrounding surface, at the highest point. Therefore, while extensive measurements and a detailed statistical analysis might reveal that the distribution of feature heights and widths present has changed, from a qualitative perspective there has been no meaningful change in the \ce{C60F48} assembly after annealing at 0.05~ML coverage. While this might suggest that a 90\celsius\, anneal is insufficient to overcome the diffusion barrier for \ce{C60F48} on hydrogen-terminated diamond, images acquired at 0.3~ML coverage following annealing demonstrate that this is not the case. In Fig.~\ref{fig:C60F48_annealed_STM}(c) it can be seen that the islands observed before annealing are no longer present, and the post-anneal surface is covered in disordered wetting layers. Through gaps present in the layer, such as Fig.~\ref{fig:C60F48_annealed_STM}(d), we are able to determine that the layer is typically a single molecular layer high, although intermittently there are molecules observed to occupy a second layer, circled in Fig.~\ref{fig:C60F48_annealed_STM}(c,d). As the wetting layers cover areas significantly larger than the islands observed before annealing, it is evident that \ce{C60F48} is mobile on the surface at 90\celsius, and upon cooling to room temperature the system is more thermodynamically stable as an extended wetting layer than as islands. The questions that this raises are, firstly, what causes this difference in behaviour with coverage, and secondly does the change in assembly at 0.3~ML cause a change in the energy level alignment? As we are unable to determine from these images what influence this annealing may have had on the relative population of charged and neutral molecules, which may be relevant to the change in assembly, we will look to answer the question of energy level alignment and the relative population of charged and neutral molecules before discussing the change in assembly in greater detail.

\subsection{X-ray Photoelectron Spectroscopy}
\label{sec:XPS}

Having established that submonolayer \ce{C60F48} forms disordered, multilayered islands upon deposition on the hydrogen-terminated diamond (100) surface and undergoes a change in assembly with annealing at higher coverage, we now wish to understand (1) the distribution of charged and neutral \ce{C60F48} within the islands which are on the surface following deposition, and (2) whether the low temperature annealing and subsequent change in assembly at 0.3~ML modifies the energy level alignment relevant to surface transfer doping. Synchrotron-based XPS measurements of the C1s core level are well suited to addressing these questions, as the charged and neutral \ce{C60F48} molecules produce components which can be resolved with careful fitting \cite{edmonds2012surface}. From the binding energy (BE) position and intensity of these components, we are able to determine the energy level alignment of \ce{C60F48} relative to the surface, and the relative amount of charged and neutral \ce{C60F48} for a given coverage \cite{edmonds2012surface}.

\begin{figure}
	\centering
	\includegraphics{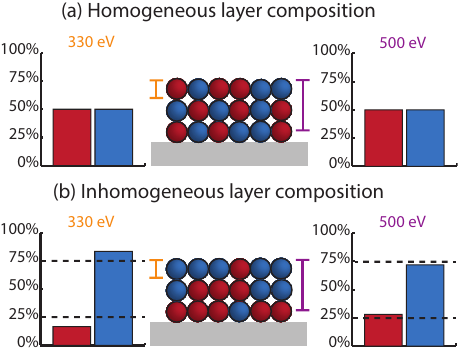}
	\caption{A sketch demonstrating the difference between the measured relative intensity of signals related to two species with a change in photon energy when islands have formed (a) a homogeneous distribution of the two species throughout the depth of the island, (b) an inhomogeneous distribution throughout the depth of the island. Arrows indicate the IMFP at each photon energy}
	\label{fig:XPS_measurement_logic}
\end{figure}

A feature of XPS is that performing measurements of the same core level with different photon energies alters the photoelectron kinetic energy (KE) of the excited core level electrons. This leads to a change in the photoelectron inelastic mean free path (IMFP) and thus the information depth of the measurement, typically regarded as three times the IMFP, allowing the depth sensitivity of XPS to be tuned. Edmonds~\textit{et al.} used a photon energy of 330~eV to characterise \ce{C60F48}-induced transfer doping of the hydrogen-terminated diamond surface, corresponding to a 3.5~\AA IMFP for C1s photoelectrons through \ce{C60F48} \cite{edmonds2012surface}. As a \ce{C60F48} molecule is approximately 10~\AA\, in height, the \ce{C60F48}-related signal in C1s core level spectra measured at 330~eV will originate almost entirely from the uppermost molecular layer in each island, and therefore the derived ratio of charged to neutral molecules is not necessarily representative of the full adlayer. To address this, in this work we have measured the C1s core level with two different photon energies, 330~eV and 500~eV; for 500~eV excitation the IMFP through the \ce{C60F48} layers is 8.4~\AA\, at 500~eV (information depth $\sim24.2$~\AA) \cite{li2006electron}. As demonstrated in Fig.~\ref{fig:XPS_measurement_logic}, a comparison of the relative intensity of the charged and neutral components extracted from these measurements, with different information depths, will allow us to qualitatively identify whether these species are homogeneously distributed through the depth of the islands.   

In order to arrive at meaningful conclusions, it is imperative that we establish a physically reasonable fitting model which will be consistently applied across the full dataset, correctly accounting for quantities which may change with the change in photon energy and IMFP, such as the intensity of specific components, while fixing quantities which should not change, such as the binding energy (BE) of specific components. Our previous work in Edmonds~\textit{et al.}, a photoelectron spectroscopy investigation of \ce{C60F48} doping of hydrogen-terminated (100) diamond performed with a photon energy of 330~eV \cite{edmonds2012surface}, provides a starting point for such a model. In the absence of surface transfer doping, \ce{C60F48} creates two C1s components in the binding energy range 285 - 290~eV, separated by $2.07\pm0.05~\rm{eV}$; the lower BE component representing \ce{C=C} bonds, and the higher representing \ce{C-F} bonds. When transfer doping is factored in, the charged and neutral molecules each produce their own pair of components, by convention labelled as \ce{(C=C)^-}/\ce{(C-F)^-} (charged) and \ce{C=C}/\ce{C-F} (neutral). These charged and neutral components are separated by $0.58\pm0.07~\rm{eV}$, with the neutral components appearing to higher binding energy. Additionally, a small but finite amount of beam-induced dissociation of \ce{C60F48} takes place during data acquisition \cite{rietwyk2011fluorination,edmonds2012surface}, requiring two more components to fit, labelled \ce{C=C_x} and \ce{C-F_x}, which do not have a fixed relationship to the other components except that they appear to lower BE of the \ce{(C=C)^-}/\ce{(C-F)^-} components. 

We have considered all of the above when developing our analysis approach. When fitting each charge state, a relative intensity of 1:4 between the \ce{C=C} and \ce{C-F} components has been maintained, as expected from the molecular stoichiometry of the molecule, in addition to the BE separations between components described above. When fitting data acquired at the same coverage and the same stage of the experiment (deposited/annealed), we have applied constraints to ensure that the charged and neutral component BE positions remain the same in the C1s core levels acquired at different photon energies, but allowed the intensities and Gaussian widths to change as required to satisfy the fit. This treatment is consistent with the logic that the \ce{C60F48} energy level alignment - reflected in the binding energy position of the components - will be unchanged by the change in photon energy and photoelectron mean free path, but the information depth of XPS increases with increasing mean free path, and so the relative amount of charged and neutral molecules may change. As the beam-damage related components do not have a fixed relationship to the other \ce{C60F48} components, and the rate of damage is sensitive to photon energy \cite{rietwyk2011fluorination}, we have allowed the beam-damage components to shift as required to produce a good fit. 

Where we must modify the approach we employed in Edmonds~\textit{et al.} is in how we fit the 283 - 285~eV region of the C1s core level, which is attributed to the hydrogen-terminated diamond (100) surface. This is because, in later work (Schenk~\textit{et al.}, Ref.~\citenum{schenk2016high}) we found inconsistencies in previously established fitting models for the hydrogen-terminated diamond (100) surface which resulted in them failing to correctly account for changes in photoelectron IMFP. Thus in this work, for fitting the part of the C1s core levels related to the diamond surface, we have employed the 3 component model developed in Schenk~\textit{et al.} specifically to account for the variation in component intensities with change in photon energy in an internally consistent fashion. This includes a component attributed to carbon in the diamond bulk (\ce{B}), a component attributed to subsurface carbon (\ce{B^*}), and a component representing the hydrogen-terminated carbon atoms at the surface (\ce{CH}). We have not imposed a constraint that \ce{B} components have matching BE positions in equivalent 330~eV and 500~eV spectra, as previous work establishes that the surface band profile resulting from transfer doping modifies the measured binding energy of the diamond bulk component due to a depth-dependent convolution which depends on the photoelectron IMFP through diamond \cite{edmonds2011surface}. However, we have required that the \ce{CH} and \ce{B^{*}} components hold a consistent BE position in 330~eV and 500~eV spectra, and that the relative intensity of the three components matches the values calculated using the approach described in Schenk~\textit{et al.} \cite{schenk2016high}. We note that, as there is a significant separation between the diamond and \ce{C60F48} related components \cite{edmonds2012surface} and we are consistently applying this approach, this modification affects a direct quantitative comparison of the diamond-related component positions with those published previously, but does not affect the trends observed or our conclusions. 

\begin{figure}
	\centering
	\includegraphics{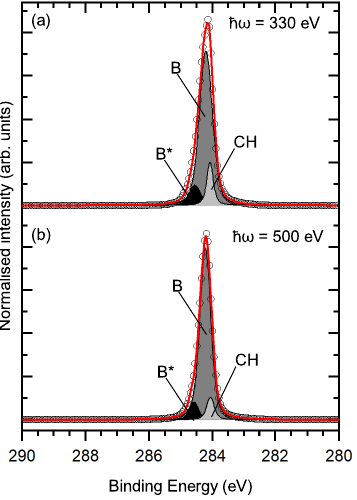}
	\caption{The C1s core level for the as-prepared hydrogen-terminated (100) diamond surface, acquired with a photon energy of (a) $\hbar\omega = 330~\rm{eV}$ and (b) $\hbar\omega = 500~\rm{eV}$.}
	\label{fig:Hterm_XPS}
\end{figure}

Figure~\ref{fig:Hterm_XPS} shows the C1s core level fits for the clean hydrogen-terminated (100) diamond surface acquired at 330~eV and 500~eV; the full details of each fit are tabulated in the supplementary material. The XPS data shows a narrow core level, and has been fitted with the three components described above; the diamond bulk \ce{B} component at a binding energy $\rm{E_{C1s, B}} = 284.22\pm0.05$~eV, with the \ce{B^*} and \ce{CH} component chemically shifted from the bulk by $+0.36\pm0.05$~eV and $-0.16\pm0.05$~eV respectively, consistent with previous work \cite{schenk2016high}. The lack of any components to higher binding energy of \ce{B^*} indicates that there is no surface oxidation \cite{kono2019carbon,dontschuk2023x} which, combined with the narrow lineshape, indicates that this is a well prepared, high quality surface.

\begin{figure}
	\centering
	\includegraphics{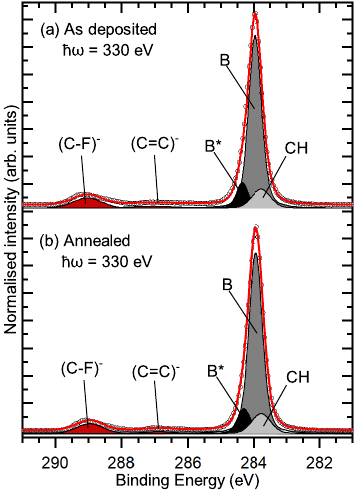}
	\caption{C1s spectra for 0.05~ML of \ce{C60F48} on hydrogen-terminated (100) diamond (a) as deposited, $\hbar\omega = 330$~eV, (b) following 90\celsius\, anneal for 10 minutes, $\hbar\omega = 330$~eV.}
	\label{fig:LC_C60F48_XPS}
\end{figure}

\begin{figure*}
	\centering
	\includegraphics{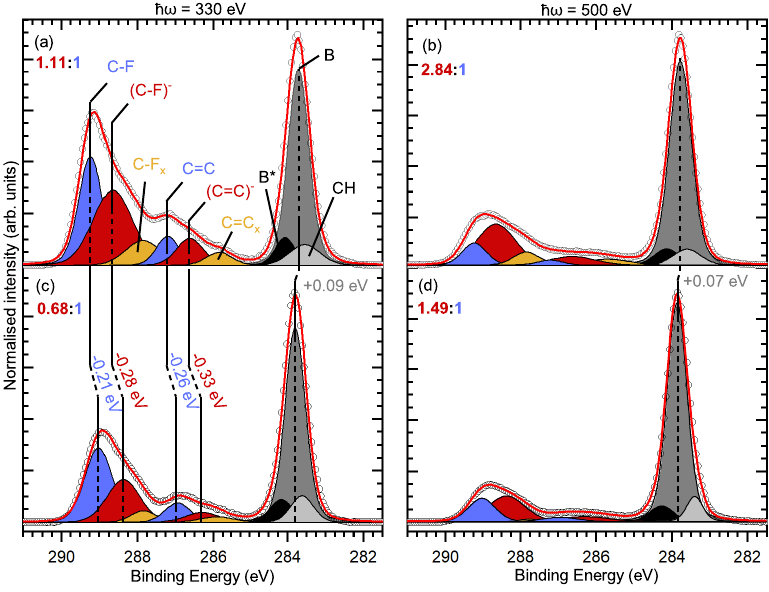}
	\caption{C1s spectra for 0.3~ML of \ce{C60F48} on hydrogen-terminated (100) diamond (a) as deposited, $\hbar\omega = 330$~eV, (b) as deposited, $\hbar\omega = 500$~eV, (c) annealed, $\hbar\omega = 330$~eV, (d) annealed, $\hbar\omega = 500$~eV. The relative shift of the components between deposition and annealing is indicated on (a,c); the same shifts apply to \ce{C60F48}-related components in (b,d), as do the component labels. The ratio of charged (red) to neutral (blue) \ce{C60F48}, calculated from the sum of the relevant \ce{C=C} and \ce{C-F} component intensities is indicated alongside each spectrum.}  
	\label{fig:HC_C60F48_XPS}
\end{figure*}

Figure~\ref{fig:LC_C60F48_XPS} shows the C1s core level acquired at a photon energy of 330~eV following deposition of 0.05~ML of \ce{C60F48}, and after subsequently annealing the 0.05~ML adlayer to 90\celsius\, for 10~minutes; the full details of each fit are tabulated in the supplementary material. The fit for the spectrum following the deposition of 0.05~ML \ce{C60F48} is as expected for low \ce{C60F48} coverage. The diamond related components are largely unchanged, except for shifting to lower binding energy ($\rm{E_{C1s, B}} = 283.96\pm0.05$~eV) as a result of transfer doping \cite{edmonds2012surface}, and all components having increased broadening as a result of inelastic scattering in the molecular adlayer. The part of the spectrum attributed to \ce{C60F48} requires only the charged \ce{(C=C)^-} and \ce{(C-F)^-} components to fit. This is consistent with prior work as, at such low coverage, all \ce{C60F48} molecules present are participating in transfer doping and therefore there are only negatively charged molecules on the surface \cite{edmonds2012surface}. There is no observable change in the spectrum following the anneal; the \ce{C60F48} peaks and diamond peaks are in the same positions, and the intensity of the \ce{C60F48} components relative to the diamond components has remained the same. This indicates that the anneal does not alter the energy level alignment or the amount and assembly of the \ce{C60F48}, consistent with the STM observation of no meaningful change at this coverage. As our fitting indicates that all molecules are charged for both sample states at this coverage, consistent with expectations \cite{edmonds2012surface}, analysis of the spectra acquired at 500~eV for each stage will not give us any further information about the observed molecular islands.

The C1s core level spectra for the 0.30~ML surface for all stages of the experiment are presented in Fig.~\ref{fig:HC_C60F48_XPS}; the full details of each fit are tabulated in the supplementary material. In the data acquired following the deposition (Figure~\ref{fig:HC_C60F48_XPS}(a,b)), the bulk diamond core level has shifted to lower BE ($\rm{E_{C1s, B}^{330}} = 283.71\pm0.05$~eV,$\rm{E_{C1s, B}^{500}} = 283.78\pm0.05$~eV) than at 0.05~ML, consistent with an increase in transfer doping. Additionally, we require both charged and neutral \ce{C60F48}-related components, as well as beam-damage components, to fit the spectra; the full details of each fit in Fig.~\ref{fig:HC_C60F48_XPS} are tabulated in the supplementary material. It is clear, even from a cursory examination, that the ratio of charged (red) to neutral (blue) \ce{C60F48} is significantly different when measured at 330~eV (1.11:1) and 500~eV (2.84:1). Given the increased information depth of the 500~eV measurement when compared to the 330~eV measurement, the relative increase in this ratio indicates that there is not an even distribution of charged molecules throughout the depth of the islands, but rather that there is a higher proportion of charged \ce{C60F48} molecules in the lower layers (closer to the surface) of the islands.

\begin{table*}
	\centering
	\caption{Component positions and calculated energy level separations at the \ce{C60F48}/diamond interface shown in Fig.~\ref{fig:C60F48_ELA}; all bulk positions are taken from the fits at 330~eV.}
	\label{tab:ELA_calcs}
	\begin{tabular}{|c|c|c|c|c|}
		\hline
		& $\rm{E_{C1s, B}} (eV)$ & $\rm{(E_F - E_{VBM}) (eV)}$ & $\rm{E_{C1s, \ce{C=C}}}$ (eV) & $\rm{(E_F - E_{LUMO}) (eV)}$\\ \hline
		0.05~ML, as deposited & 283.96 & +0.06 & 287.05\footnote{Component position calculated from the position of the \ce{(C=C)^-} component, assuming a 0.58~eV separation} & -0.43\\ \hline
		0.30~ML, as deposited & 283.71 & -0.19 & 287.21 & -0.24 \\ \hline
		0.30~ML, annealed & 283.80 & -0.10 & 286.95 & +0.11 \\ \hline		 
	\end{tabular}
\end{table*}

In the spectra acquired following the 90\celsius\, anneal (Fig.~\ref{fig:HC_C60F48_XPS}(c,d)) the \ce{C60F48} and diamond components have shifted, and the intensity of the \ce{C60F48} components has changed in both the 330~eV and 500~eV spectra. The change in the intensity of the \ce{C60F48} components relative to the diamond components is unsurprising given that the change in film morphology will, for surface sensitive XPS measurements, significantly impact the attenuation which contributes to the relative size of these components. However, the charged to neutral ratio decreasing is a change which needs careful consideration. We know that at this coverage the 90\celsius\, anneal will cause a ``flattening'' of the islands and result in wetting layers. Our comparison of the 330~eV and 500~eV data following deposition suggests that the bottom layers of the islands contain a higher proportion of charged molecules than the upper layers, and so we might have anticipated seeing an increase in this ratio, if any change at all. A decrease in this ratio suggests that there has been a change in the amount of charge transferred to the adlayer from the substrate as a result of the anneal. The shift of the diamond bulk towards higher BE (an average $+0.08\pm0.05$~eV) supports this, as an upward shift indicates a decrease in the carrier density of the 2DHG \cite{edmonds2012surface}. This is accompanied by the \ce{C60F48} components shifting to lower BE by an average $-0.27\pm0.05$~eV, indicating a change in the energy level alignment between the surface and the adlayer. From the measured diamond bulk ($\rm{E_{C1s, B}}$) and \ce{C60F48} \ce{C=C} component ($\rm{E_{C1s, \ce{C=C}}}$) positions, as well as the known separations between the diamond bulk component and VBM ($283.9\pm0.1$~eV) \cite{maier2001electron} and the \ce{C60F48} \ce{C=C} component and LUMO ($287.16$~eV) \cite{edmonds2012surface}, we are able to determine the energy level alignment at the diamond/\ce{C60F48} interface by making use of the equations:

\begin{equation}
	\rm{(E_F - E_{VBM}) = E_{C1s, B} - 283.9} 
\end{equation}
\begin{equation}
	\rm{(E_F - E_{LUMO}) = E_{C1s, \ce{C=C}} - 287.16}
\end{equation}

\begin{figure}
	\centering
	\includegraphics[width=82mm]{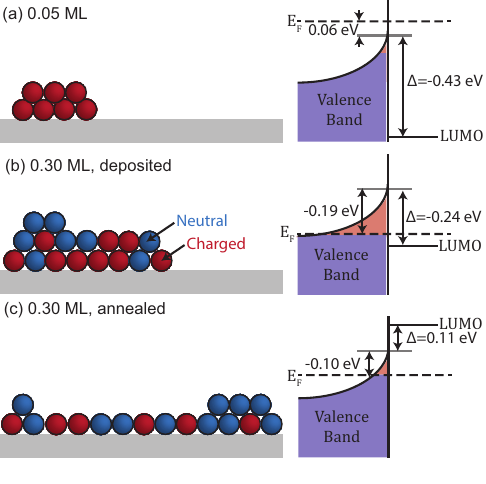}
	\caption{A sketch of the \ce{C60F48} assembly and the energy level alignment at the \ce{C60F48}/diamond interface derived from XPS measurements performed at $\hbar\omega = 330$~eV for (a) 0.05~ML, (b) 0.30~ML as deposited, (c) 0.30~ML following the 90\celsius\, anneal.}
	\label{fig:C60F48_ELA}
\end{figure}

Fig.~\ref{fig:C60F48_ELA} shows the alignment for all stages of this experiment calculated from these equations, using the component positions in Table~\ref{tab:ELA_calcs}. It is apparent that the energy level alignment has changed substantially in response to the change in assembly, with the \ce{C60F48} acceptor energy changing by $+0.35\pm0.10$~eV, rendering it positive and thereby indicating that charge transfer from the surface into the \ce{C60F48} adlayer is energetically unfavourable in this state. This inference is consistent with the observed increase in the relative proportion of neutral \ce{C60F48}, as such a change in alignment would allow charge to transfer from the adlayer into the diamond surface, reducing the charge in the adlayer, until a new equilibrium was reached. From the measured doping efficiencies we are able to estimate the amount of charge in the adlayer, and therefore the 2DHG carrier density, following the deposition and the 90\celsius\, anneal.  Assuming that 1~ML of \ce{C60F48} corresponds to a molecular surface density of $7.8\times10^{13}~\rm{molecules/cm^{2}}$ \cite{edmonds2012surface}, the doping efficiency determined from the charged to uncharged molecule ratios measured for a 0.3~ML \ce{C60F48} coverage in the 330~eV (500~eV) data gives a carrier density of $1.2\times10^{13}~\rm{cm^{-2}}$ ($1.7\times10^{13}~\rm{cm^{-2}}$) following deposition, and $9.5\times10^{12}~\rm{cm^{-2}}$ ($1.4\times10^{13}~\rm{cm^{-2}}$) following the anneal. It is not possible to determine more precise values of the carrier density from photoemission measurements without extensive modelling of the molecular assembly and photoelectron attenuation. However, what these values do demonstrate is that the carrier density decreases by approximately 20\% in response to the change in the molecular assembly at the surface and the accompanying change in energy level alignment. This is a substantial decrease in an applied context, critically for spintronic devices where the strength of the spin-orbit coupling is strongly dependent on the carrier density \cite{akhgar2016strong}.  

\subsection{Submonolayer assembly of \ce{C60F48} at the hydrogen-terminated (100) diamond surface}
\begin{figure*}
	\centering
	\includegraphics[width=172mm]{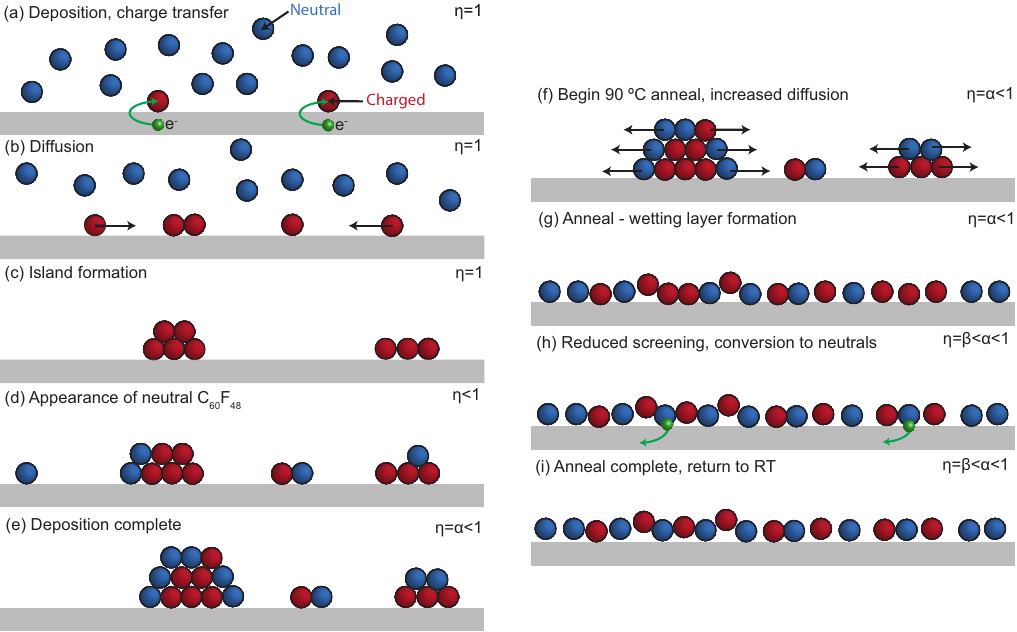}
	\caption{A schematic diagram of the proposed processes occurring during (a) - (e) the deposition of \ce{C60F48}, and (f) - (i) annealing the \ce{C60F48} adlayer, panels each described in text.}
	\label{fig:C60F48_assembly}
\end{figure*}

Having observed a change in both the assembly and the interfacial energy level alignment with 90\celsius\, annealing at 0.3~ML, but not at 0.05~ML, let us now attempt to explain what drives this behaviour. It is clear from our STM measurements at 0.3~ML that 90\celsius\, is sufficiently high temperature to overcome the surface diffusion barrier for \ce{C60F48} on the hydrogen-terminated diamond (100) surface. We infer from this that \ce{C60F48} does diffuse across the surface at 0.05~ML when annealed, but that it is thermodynamically favourable for the dewetting islands to reform when the system cools to room temperature. That the energy level alignment at 0.05~ML (Fig.~\ref{fig:C60F48_ELA}(a)) guarantees that all \ce{C60F48} molecules will be charged at this coverage, while there are neutral molecules present at 0.3~ML, indicates that the presence of neutral molecules is a relevant factor in the wetting layer formation. This, rather naturally, points towards it being the competing repulsive (electrostatic) and attractive (van der Waals) interactions which give rise to this behaviour, as neutral molecules will participate in the latter but not the former. Given the short-range nature of the van der Waals interaction, the dewetting island reformation at 0.05~ML can therefore be understood as simply being the most effective configuration to maximise the number of nearest neighbours, and therefore the number of stabilising van der Waals interactions each molecule is participating in. 

To understand why we see island formation following deposition of 0.3~ML, but not island reformation following the anneal, we must consider the interplay of transfer doping and assembly during these processes. Importantly, we must recognise that initially all molecules which are deposited become charged, and it is only in later stages of deposition that neutral molecules appear, while the anneal begins with a mixture of charged and neutral molecules. This difference can be expected to influence the evolution of the assembly, as neutral molecules offer a means to increase the number of stabilising van der Waals interactions without a competing repulsive interaction. Figure~\ref{fig:C60F48_assembly} presents a proposed scheme for the assembly during deposition and the annealing process. Given the unity doping efficiency, the first \ce{C60F48} molecules which deposit onto the surface become charged (a) and, as they diffuse across the surface, encountering other molecules and forming small islands (b), which then grow larger (c) as other molecules diffuse and encounter these islands. These three processes (a) - (c) proceed in parallel with only charged molecules contributing, until the coverage reaches a point where the doping efficiency drops below 1, at which point neutral molecules begin to participate in these processes (d), joining islands that have been nucleated by charged \ce{C60F48} molecules or nucleating their own islands. At the end of the deposition, the surface is populated with multilayered islands. This then serves as the starting point when the 90\celsius\, anneal is started (f), which increases the diffusion rate of the \ce{C60F48}. The presence of neutral molecules is key to the behaviour in this stage - rather than molecules diffusing and having a thermodynamic drive to form islands, the attractive van der Waals interactions offered by the neutral molecules, without also having repulsive interactions with other molecules, allows \ce{C60F48} to form an extended layer (g). The transition from islands to an extended layer changes the dielectric screening of the \ce{C60F48} making up the layer, affecting the energy level alignment between the acceptors and the surface and the characteristic capacitance $\rm{C_\square}$, and so some amount of charge is transferred back to the diamond surface (h), thereby reducing the doping efficiency of the adlayer and reducing the electrostatic potential felt by each charged molecule in the adlayer, further increasing the thermodynamic stability of the layer arrangement such that it remains once the layer is cooled (i). We are unable to definitively demonstrate that this proposed model is what occurs without extensive further work, but note that it qualitatively fits with our experimental observations and the current understanding of surface transfer doping and molecular assembly

Finally, we consider the broader impact our findings may have for future studies of molecule-induced surface transfer doping and the translation of this research to practical devices. In recent years photoelectron spectroscopy has emerged as a useful tool for studying molecular transfer doping, due to the ability to directly measure the relative population of charged and uncharged molecules, and to observe changes in the Fermi level position and key aspects of the band alignment \cite{edmonds2012surface,sear2017p}. Such investigations typically use low photon energies which yield close to the minimum IMFP in order to maximise sensitivity to the species in the adlayer, as well as benefit from the improved energy resolution that this typically allows. However, the data analysis in such studies has typically relied on the assumption that the at submonolayer coverages the film grows in a layer-by-layer fashion. While this may be the case for some acceptors, our results demonstrate that this is not always the case, potentially affecting an accurate determination of the proportion of charged and neutral molecules making up the adlayer. While a quantitatively significant microscopy study to evaluate the assembly of a given acceptor may be infeasible, a qualitative assessment can be made by performing a sequence of photoelectron spectroscopy measurements at photon energies corresponding to sufficiently different IMFP values that the information depth of the core level meaningfully changes, as we have done here. Furthermore, we have demonstrated that the assembly of molecular acceptors can meaningfully influence the energy level alignment between the surface and the adlayer and therefore, by extension, the charge transfer doping process. While other molecular acceptors may not undergo such a transition at as low a temperature as \ce{C60F48}, or have as significant a change in the energy level alignment as a consequence, it will be important to evaluate this for molecular acceptors which are proposed as ``thermally stable'' dopants, given that it is clearly possible for a change in assembly to have significant bearing on the transfer doping process. Beyond the change in the 2DHG which we can infer from this work, it is also reasonable to expect that the change in the spatial distribution of ionized molecular acceptors will impact the transport mobility of the 2DHG \cite{daligou20202d}, and influence the localization behaviour that governs carrier transport at low temperature \cite{xing2020strong}. 

\section{Conclusion}
We have studied the submonolayer assembly of \ce{C60F48} on the hydrogen-terminated diamond (100) surface using a combination of STM and XPS, the first published work which characterises the influence of molecular assembly on the energy level alignment in a transfer-doped diamond system. We find that upon deposition \ce{C60F48} forms disordered, multilayered islands on the surface, rather than the layer-by-layer growth mode assumed in previous work, an arrangement which provides an explanation for the anomalously large electron affinity for \ce{C60F48} on the hydrogen-terminated diamond surface determined in previous work. This Volmer-Weber growth is likely the result of a weak surface-molecule interaction, a low surface diffusion rate at room temperature, \ce{C60F48} screening the molecular charge and a strong attractive intermolecular van der Waals force. Following annealing to 90\celsius\, there is a transition to dewetting layers at higher coverage, accompanied by a change in the energy level alignment relevant to tranfer doping and a decrease in the amount of charge transferred to the adlayer. These observations demonstrate that while molecular acceptors may be resistant to decomposition and desorption at higher temperatures, there is still a need to consider whether elevated temperatures may cause a change in molecular assembly and the energy level alignment relevant to surface transfer doping. 

\section*{Acknowledgements}
This research was undertaken using the Soft X-ray Spectroscopy beamline at the Australian Synchrotron, part of ANSTO. The authors gratefully acknowledge discussions with Lothar Ley. 

\bibliography{Molecular_assembly_bib}

\end{document}